# POWER MANAGEMENT DURING SCAN BASED SEQUENTIAL CIRCUIT TESTING

Reshma.p

M.tech Scholar, Department of Electronics and Communication Engineering, Amrita Vishwa Vidyapeetham University, Coimbatore.
`reshma_p2004@rediffmail.com`

*ABSTRACT*

*This paper shows that not every scan cell contributes equally to the power consumption during scan based test. The transitions at some scan cells cause more toggles at the internal signal lines of a circuit than the transitions at other scan cells. Hence the transitions at these scan cells have a larger impact on the power consumption during test application. These scan cells are called power sensitive scan cells.A verilog based approach is proposed to identify a set of power sensitive scan cells. Additional hardware is added to freeze the outputs of power sensitive scan cells during scan shifting in order to reduce the shift power consumption.when multiple scan chain is incorporated along with freezing the power sensitive scan cell,over all power during testing can be reduced to a larger extend.*

*KEYWORDS*

*Scan ,Scan DFlip-Flop ,DFTAdvisor*

## 1. INTRODUCTION

With current advances in very large scale integrated technology , the sensitivity of today's chips to deep submicrometer effects is increasing. Along with technology scaling, the increase in the operating frequency and the increase in the functional density of today's digital designs has led to new challenges for designers and test engineers. Furthermore, dynamic power consumption and IR-drop due to excessive switching activity are critical challenges. As a result, power reduction techniques have been extensively studied by both industry and academia with respect to both design and test. Scan-based test remains one of the most widely accepted design-for-test techniques because it significantly improves the controllability and the observability of the circuit's internal nodes with an insignificant area and performance overhead.Switching activity during scan-based test is often much higher than that during normal operation. There are multiple reasons for this phenomenon. First, the test vectors applied consecutively are not correlated. Second, nonfunctional states may be traversed during scan-test. Furthermore, test compaction and testing multiple cores simultaneously contribute toward high-switching activity. In addition, as patterns are shifted into and out of the scan chains, multiple changes of the flip-flop values can propagate into the combinational logic and cause massive amounts of switching.

Scan-based tests might cause excessive circuit switching activity compared to a circuit's normal operation . Higher switching activity causes higher peak supply currents and higher power dissipation. High power dissipation during test can cause many problems, which are generally addressed in terms of average power and peak power. Average power is the total distribution of power over a time period, and is calculated using the ratio of consumed energy to test time .Peak power is the highest power value at any given instant. When peak power is beyond the design limit, a chip cannot be guaranteed to function properly due to additional gate delays caused by the supply voltage drop. The power consumption within one clock cycle may not be large enough to elevate the temperature over the chip's thermal capacity limit. To damage the chip, high power consumption must last for an enough number of clock cycles.The test power consumed during scan shifting and capture cycles is referred to as shift power and capture power, respectively. A typical scan chain in industrial designs consists of at least hundreds of scan cells, whereas the capture window only lasts one or a few clock cycles. Clearly, the average power consumption is determined by the shift power. Excessive shift power accumulation may make a good chip fail during test even if the peak capture power is low. Inserting no operation cycles between the end of scan shifting and the beginning of capture cycles can reduce the chance of rejecting good chips during test. This experimental result implies that test power reduction strategies should focus on reduction of the average shift power and the peak capture power.This paper, focus on reducing the average shift power in scan-based tests. An effective method to identify a set of power sensitive scan cells is used. By inserting additional hardware to freeze power sensitive scan cell outputs to pre-selected values, the average shift power consumption can be reduced dramatically.

The paper is organized as follows. In Section 2,previous work. In Section 3,proposed method. In Section 4, experimental results on S27 is presented, followed by conclusions in Section 5.

## 2.The Previous work

To reduce the switching activity during scan shift, automatic test pattern generation (ATPG)-based approach and 2) DFT-based approach were used. The advantage of the ATPG based solutions is that they do not modify the original design and the scan architecture,but modification is done on test vectors and there by power reduction can be obtained. DFT-based solutions require one to either partition the conventional scan chain architecture or insert additional hardware into the design .Different DFT approach used here are included in this literature survey. In Minimised power consumption for scan based Bist, extra logics are inserted to hold the outputs of all the scan cells at constant values during scan shifting. This methods not only minimize the average scan shift power, but also avoid peak power hazards during scan shifting. The main disadvantage of these approaches is the large area overhead, since additional logics are added to all the scan cells. Moreover, it may degrade circuit performance due to extra logics added between scan cell outputs and functional logics. To reduce the area overhead due to additional gates, supply gating transistors for the first-level gates at the outputs of scan cells are proposed in Low power scan design using first level supply gating. An alternative implementation to hold the scan cell outputs by using dynamic logic was proposed in Techniques for minimizing power dissipiation in scan and combinational circuit during test application.The method proposed in Inserting test points to control peak power during scan testing, inserts test points at selected scan cell outputs to keep the peak shift power at every shift cycle below a specified limit. Given a set of test patterns, logic simulation is carried out to identify the shift cycles in which peak power violations occur. Those cycles are called violating cycles. By using integer linear programming (ILP) techniques, the optimization problem is solved to select as few test points as possible such that all violating cycles can be eliminated. In Partial gating optimization for power reduction during test application, random vector simulation was used to guide partial test point selection. When simulating a random vector, the primary inputs and the pseudo primary inputs are changed to value X with pre-specified probabilities, and the number of gates becoming X after the change is used as a cost function to identify the logic value assigned at the primary inputs and the pseudo primary inputs, as well as to select scan cells to be held during scan shifting. To explore several hundred thousands of scan cells in an industrial circuit, a significant number of random vectors need to be simulated in order to choose good test points.

Motivated by the test point insertion approach along with multiple scan chain, scan shift power can be reduced to a larger extend. Some scan cells have a much larger impact on toggle rates at the internal signal lines than other scan cells. These scan cells are called power sensitive scan cells. Objective is to quickly identify power sensitive scan cells and their preferred frozen values during scan shifting. By freezing a small percentage of scan cells that are the most power sensitive, reduction in scan shift power can be achieved ,while minimizing the additional area overhead. Compared with the previous approaches , this approach has less area overhead and can avoid modifying scan cells at critical paths by not selecting them to freeze. This approach also provides a practical way to handle large industrial designs and since both freezing power sensitive scan cells and multiple scan chain approach is used, power can be reduced to a larger extend.

## 3.The Proposed Method

A mentorgraphics tool called DFTAdvisor is used for Scan insertion.An S27 ISCAS benchmark circuit is taken and scan is inserted in that circuit using DFTAdvisor.By using Fastscan all the testvectors for stuck at faults are identified.By simulating the testbench obtained from Fastscan,along with the verilog code obtained from the DFTAdvisor,number of toggling are calculated.By using a counter toggling are counted and identify the point where the logic gate has to be inserted.In S27 circuit it is find that the toggling in between the first and zeroth ScanD-FF is higher.By introducing logic gates at the output of first ScanD-FF we can block the complete toggling occurring at the combinational part,when only the Flip-Flops are selected. An additional AND gate with the second input inverted is inserted between the scan cell output and the function logics it drives. During scan shifting, the scan enable signal se is asserted to be 1. It makes additional gate output become constant. During normal operations, se is de-asserted to be 0 and the additional gates do not impact normal operations.

## 4. Experimental Results

For testing a sequential circuit, both combinational cloud testing as well as memory element testing is needed. The approach used here converts all D-flip-flop to scanD-flipflop.DFTAdvisor is a synthesis tool capable of doing scan insertion. It accepts gate level net list format and generates a new net list with scan cells inserted.The circuit in Fig.1 shows ISCAS S27 benchmark circuit.when the synthesisable netlist code for it is given to DFTAdvisor,it will generate a code with scan inserted as shown in Fig.2. When the code obtained from DFTAdvisor is given to Fastscan,corresponding test vectors obtained are as shown in Fig.3.By simulating the testbench obtained from Fastscan,along with the code from DFTAdvisor, the result is as shown in Fig.4

Fig.1 ISCAS S27 Benchmark Ckt

Fig.2a. Invocation of DFTAdvisor

Fig.2b. Scan inserted ckt obtained from DFTVisualiser

```
/*
*   DESC: Generated by DFTAdvisor at Sat Dec 4
14:22:11 2010
*/
module s27 ( CLK , G0 , G1 , G17 , G2 , G3 , scan_in1 ,
scan_out1, scan_en );
input CLK , G0 , G1 , G2 , G3 , scan_in1 , scan_en ;
output G17 , scan_out1 ;
wire G9 , G16 , G15 , G12 , G8 , G14 , G7 , G13 , G6 , G11 ,
G5 G10 ;
wire [3:0] \$dummy ;
sff reg_d_out_0 (.D( G10 ) , .SI( \$dummy [1] ) , .SE(
scan_en ) , .CLK ( CLK ) , .Q( G5 ) , .QB ( \$dummy [0]
));
sff reg_d_out_1 (.D( G11 ) , .SI( \$dummy [2] ) , .SE(
scan_en ) , .CLK ( CLK ) , .Q( G6 ) , .QB ( \$dummy [1]
));
sff reg_d_out_2 (.D( G13 ) , .SI( scan_in1 ) , .SE(
scan_en ) , .CLK ( CLK ) , .Q( G7 ) , .QB ( \$dummy [2]
));
inv02 ix249 (.A ( G0 ) , .Y ( G14 ));
inv02 ix250 (.A ( G11 ) , .Y ( G17 ));
and02 ix155 (.A0( G6 ) , .A1( G14 ) , .Y ( G8 ));
or02 ix211 (.A0( G8 ) , .A1( G12 ) , .Y ( G15 ));
or02 ix212 (.A0( G3 ) , .A1( G8 ) , .Y ( G16 ));
nand02 ix157 (.A0( G15 ) , .A1( G16 ) , .Y ( G9 ));
nor02 ix215 (.A0( G14 ) , .A1( G11 ) , .Y ( G10 ));
nor02 ix216 (.A0( G5 ) , .A1( G9 ) , .Y ( G11 ));
nor02 ix217 (.A0( G7 ) , .A1( G1 ) , .Y ( G12 ));
nor02 ix218 (.A0( G2 ) , .A1( G12 ) , .Y ( G13 ));
assign scan_out1 = \$dummy [0] ;
endmodule
```

Fig.2c.Scan inserted code generated by DFTAdvisor

```
// Tessent FastScan v9.2
// Design = dfta_out/s27_scan.v
// Created = Fri Jan 14 06:31:46 2011
//    Test Coverage  = 100.00%
//    Total Faults   = 110
//    Total    Patterns = 8
//    Fault Type   = stuck
ASCII_PATTERN_FILE_VERSION = 2;
SETUP=
  declare input bus "PI" = "/CLK","/G0","/G1",
"/G2","/G3","/scan_in1","/scan_en";
  declare output bus "PO" = "/G17","/scan_out1"
CHAIN_TEST=
pattern = 0;
  apply "grp1_load" 0 =
    chain "chain1" = "010";
  end;
  force  "PI" "0XXXXXX" 1;
  measure "PO" "XX" 2;
  apply "grp1_unload" 3 =
    chain "chain1" = "101";
  end;
```

Fig.3.Test vectors obtained from Fastscan

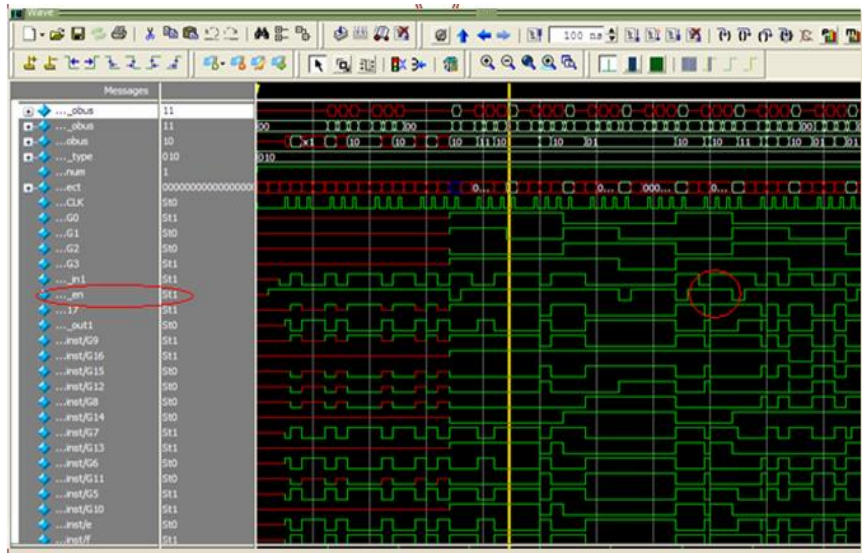

Fig.4.Simulated result

From the simulated result, it is clear that when scan enable is high there will be toggling at combinational part.But it is not desired.when scan enable is high ther will not be any toggling at combinational part,only flip-flop is involved.By using a counter,the toggligs when scan enable is high are calculated and found that toggling between Dff_1 and Dff_0 , and Dff_2 and Dff_0 are higher.By introducing logic gates at these points it is seen that overall toggling can be reduced to a larger extend.It is shown in Fig.5.

| Flip Flops | Gates Involved | Toggling |
|---|---|---|
| Dff_1 – Dff_0 | G8 ,G16,G15,G9,G11,G10 | 88 |
| Dff_2 - Dff_0 | G12,G15,G9,G11,G10 | 88 |
| Dff_2- Dff_1 | G12,G15,G9,G11 | 82 |

Fig.5.a.Toggling after scan insertion

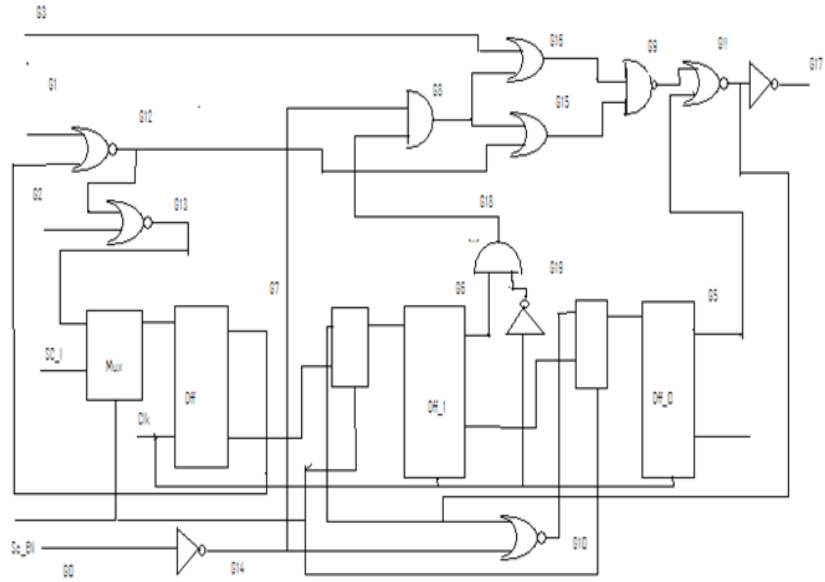

Fig.5.b.Logic insertion between Dff_1 and Dff_0

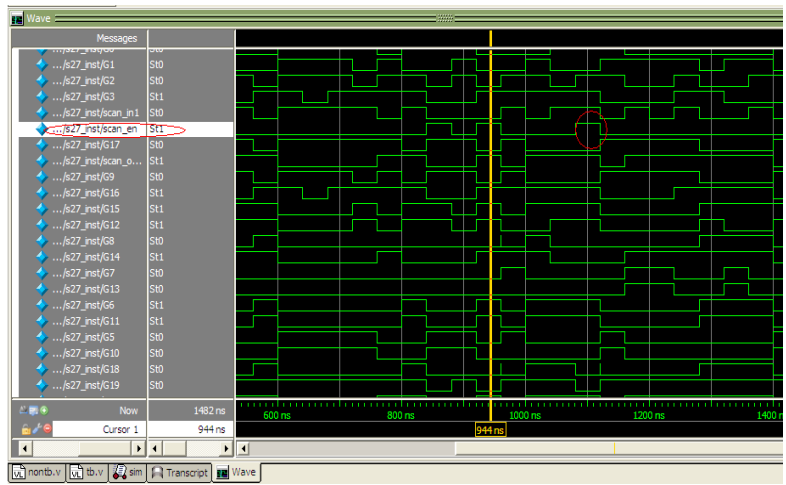

Fig.5.c.Simulation result after gate insertion

| Flip Flops | Gates Involved | Toggling |
|---|---|---|
| Dff_1 – Dff_0 | G8, G16, G15, G9, G11, G10, G18 | 38 |
| Dff_2 - Dff_0 | G12, G15, G9, G11, G10 | 36 |
| Dff_2 - Dff_1 | G12, G15, G9, G11 | 27 |

Fig.5.d.Toggling after logical insertion

Area will be a trade-off for power.By simulating S27 in XILINX, it is observed that will not be that much increase in area overhead.It is shown in Fig.6.By using multiple scan chains ,it is verified that time required drastically decreases.corresponding simulation results are shown in Fig .7.

| 781 Project Status | | | |
|---|---|---|---|
| Project File: | 781.ise | Current State: | Placed and Routed |
| Module Name: | s27 | • Errors: | No Errors |
| Target Device: | xcv50e-6cs144 | • Warnings: | 6 Warnings |
| Product Version: | ISE, 8.1i | • Updated: | Thu Mar 24 08:08:53 2011 |

| Device Utilization Summary | | | | |
|---|---|---|---|---|
| Logic Utilization | Used | Available | Utilization | Note(s) |
| Number of 4 input LUTs | 2 | 1,536 | 1% | |
| Logic Distribution | | | | |
| Number of occupied Slices | 1 | 768 | 1% | |
| Number of Slices containing only related logic | 1 | 1 | 100% | |
| Number of Slices containing unrelated logic | 0 | 1 | 0% | |
| Total Number of 4 input LUTs | 2 | 1,536 | 1% | |
| Number of bonded IOBs | 5 | 94 | 5% | |
| Total equivalent gate count for design | 12 | | | |
| Additional JTAG gate count for IOBs | 240 | | | |

Fig.6.a.Utilization of area in S27

| RESHM Project Status | | | |
|---|---|---|---|
| Project File: | reshm.ise | Current State: | Placed and Routed |
| Module Name: | s27 | • Errors: | No Errors |
| Target Device: | xcv50e-6cs144 | • Warnings: | 4 Warnings |
| Product Version: | ISE, 8.1i | • Updated: | Thu Mar 24 08:17:04 2011 |

| Device Utilization Summary | | | | |
|---|---|---|---|---|
| Logic Utilization | Used | Available | Utilization | Note(s) |
| Number of 4 input LUTs | 8 | 1,536 | 1% | |
| Logic Distribution | | | | |
| Number of occupied Slices | 4 | 768 | 1% | |
| Number of Slices containing only related logic | 4 | 4 | 100% | |
| Number of Slices containing unrelated logic | 0 | 4 | 0% | |
| Total Number of 4 input LUTs | 8 | 1,536 | 1% | |
| Number of bonded IOBs | 8 | 94 | 8% | |
| Total equivalent gate count for design | 48 | | | |
| Additional JTAG gate count for IOBs | 384 | | | |

Fig.6.b.Utilization of area in S27 after scan insertion

| INSERTED1 Project Status | | | |
|---|---|---|---|
| Project File: | inserted1.ise | Current State: | Placed and Routed |
| Module Name: | s27 | • Errors: | No Errors |
| Target Device: | xcv50e-6cs144 | • Warnings: | 4 Warnings |
| Product Version: | ISE, 8.1i | • Updated: | Thu Mar 24 08:23:37 2011 |

| Device Utilization Summary | | | | |
|---|---|---|---|---|
| Logic Utilization | Used | Available | Utilization | Note(s) |
| Number of 4 input LUTs | 5 | 1,536 | 1% | |
| Logic Distribution | | | | |
| Number of occupied Slices | 3 | 768 | 1% | |
| Number of Slices containing only related logic | 3 | 3 | 100% | |
| Number of Slices containing unrelated logic | 0 | 3 | 0% | |
| Total Number of 4 input LUTs | 5 | 1,536 | 1% | |
| Number of bonded IOBs | 8 | 94 | 8% | |
| Total equivalent gate count for design | 30 | | | |
| Additional JTAG gate count for IOBs | 384 | | | |

Fig .6.c.Utilization of area in S27 after logical insertion between Dff_1 and Dff_0

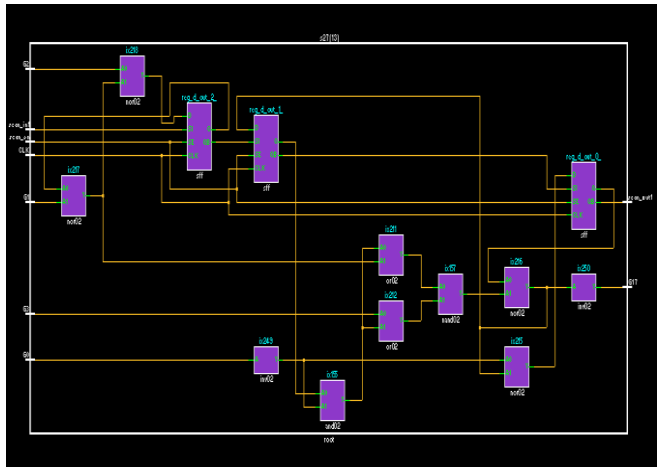

Fig .7.a.   1_scan chain

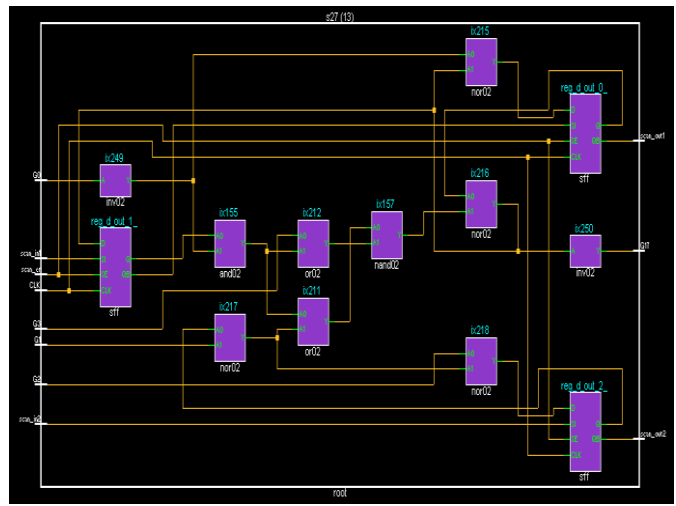

Fig .7.b.   2_scan chain

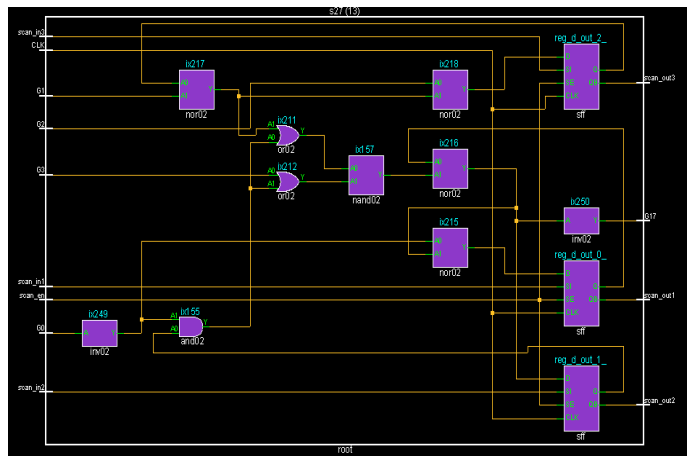

Fig .7.c.   3_scan chain

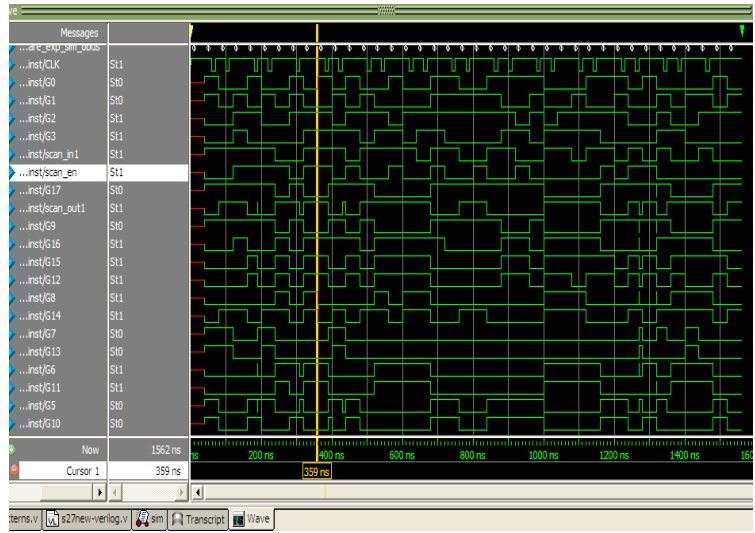

Fig .7.d.   Simulation result_1 scan chain

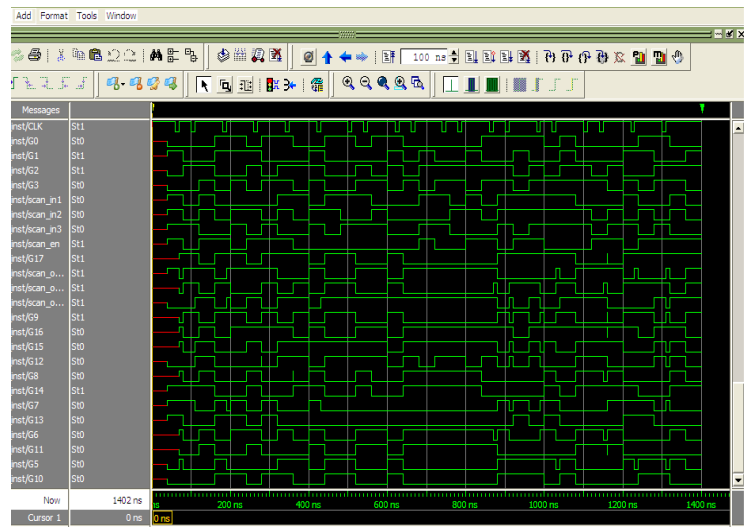

Fig .7.e.   Simulation result-3 Scan chain

|  | 1 Scan chain | 3 Scan chain |
| --- | --- | --- |
| Number of Clock required | 24 | 20 |
| Time required | 1600ns | 1400ns |

Fig .7.f.   Comparison result

# 5. CONCLUSION

This paper presented and analyzed a method for reducing switching activity during scan shift by freezing a small subset of all flip-flops at the RTL. Large reductions in switching activity can be achieved with very low-area overhead. The amount of scan flip-flops that are going to be frozen can be decreased/increased depending on the design's overhead budget. In comparison with previous methods, which freeze these flip-flops at the gate level, timing closure can be more easily met. When flip-flops are frozen at the gate level, individual timing analysis have to be implemented to determine whether or not each flipflop could be frozen without violating timing. By freezing all flip-flops simultaneously at the RTL,this approach allow the synthesis tool to automatically optimize for timing closure.

In addition, this paper has presented a detailed analysis of the switching activity reduction that can be obtained with very few frozen flip-flops. In fact, in one case, a 79% reduction in switching activity was achieved with an area overhead of only 0.02%. This switching activity analysis considered both hazards and final circuit values.when multiple scan chain is applied along with freezing powersensitive scan cells, the test power is seen to reduced to a larger extend along with that time required for testing also get reduced.